\documentstyle[twoside,fleqn,espcrc2,epsf]{article}

\title{{\normalsize
HUPD-9616~~~~%
FERMILAB-CONF-96/281-T~~~~%
\hfill hep-lat/9609014~~~~%
}\\[12pt]
The Light Quark Masses with an $O(a)$-Improved Action}
\author{T. Onogi\address{Dept.~of Physics, Hiroshima Univ.,
        1-3-1 Kagamiyama, Higashi-Hiroshima, Hiroshima 739, Japan}
        \thanks{Presented by T.Onogi.},
	A.X. El-Khadra\address{Dept.~of Phys,
        Univ.~of Illinois, 
        1110 W. Green St. Urbana, IL 61801, U.S.A.},
        B.J. Gough\address{T-8, 
        Los Alamos National Laboratory, 
        Los Alamos, NM 87545, U.S.A.},
        G.M. Hockney\address{Theory Group, Fermilab,
        P.O.Box 500, Batavia, IL 60510, U.S.A.}
        A.S. Kronfeld$^{\rm d}$,
        P.B. Mackenzie$^{\rm d}$,
        B.P. Mertens\address{Dept.~of Phys., Univ.~of Chicago,
        5640 Ellis Ave. Chicago, IL 60637, U.S.A.},
        J.N. Simone$^{\rm d}$}
\begin{document}
\begin{abstract}
We present the recent Fermilab calculations of the masses of the light
quarks, using tadpole-improved Sheikholeslami-Wohlert (SW) quarks.
Various sources of systematic errors are studied. Our final result for
the average light quark mass in the quenched approximation evaluated
in the $\overline{MS}$ scheme is $\overline{m}_q(\mu=2\,\mbox{GeV};n_f=0)=
(m_u+m_d)/2=3.6 \pm 0.6$ MeV.
\end{abstract}
\maketitle
\section{Introduction}
We present recent results on the light quark mass determination using the
SW action \cite{SW85}, which are updates of the last year's results
\cite{GOS95}. For results from wilson and staggered fermions see
\cite{Uka93}\cite{A94}\cite{GB96}.

The basic procedure is to extract the pseudoscalar masses ($m_{PS}$)
numerically for a range of quark masses and determine the linear
coefficient in the chiral extrapolation,
\begin{equation}
(m_{PS} \, a )^2 = A \, \widetilde{m}_q^{lat} \, a
\end{equation}
where $\widetilde{m}_q^{lat} = \ln(1 +
1/2\tilde{\kappa}-1/2\tilde{\kappa}_c)$, with $\tilde{\kappa} \equiv
\kappa u_0$ and $u_0 \equiv \sqrt[4]{<U_P>_{MC}}$ \cite{LM93}.

Using the experimentally measured pion mass as an input, we obtain the
light quark bare mass $\widetilde{m}_q^{lat}$, which is the average of the
up and down quark masses. We convert it to the light quark mass
$\overline{m}_q$in the $\overline{MS}$ scheme by perturbation theory.
%

%
%
Table~\ref{tab:lattices} shows the lattices used for the simulation.
We use the SW fermion action. For $\beta$\,=\,5.5, 5.7 and 5.9 the
clover coefficient $c$ is the tadpole improved tree-level value
$1/u_0^3$. However, for $\beta$\,=\,6.1, we use $c$\,=\,1.40 instead of
$1/u_0^3$\,=\,1.46. All calculations are done in the quenched
approximation. The lattice spacing $a$ is determined from the 1P--1S
charmonium splitting.
%
\begin{table}
\caption{Lattice details ($n_f=0$)}
\label{tab:lattices}
\begin{center}
\begin{tabular}{l|cccc}
\hline
\hline
$\beta$	&	5.5 & 	5.7	&	5.9&	6.1 \\
\hline
$\kappa$'s &4	& 4	& 4	& 4  \\
configs    &40	& 300	& 100	& 100  \\
$a^{-1}$\ (GeV) &0.79	& 1.16	& 1.80	& 2.55  \\
$L^3$ $(T=2L)$ & $8^3$	& $12^3$	& $16^3$	& $24^3$  \\
$L^{phys}$ (fm)   & 2.0		& 2.1		& 1.8		& 1.9  \\
$c$  & 1.69 	& 1.57	& 1.50	& 1.40  \\
\hline
\hline
\end{tabular}
\end{center}
\end{table}
%
\section{Systematic Errors}
\begin{figure}[t]
\centerline{\epsfysize=5.0cm \epsffile{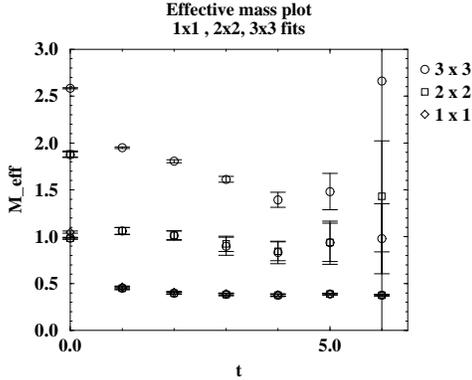}}
\caption{The pseudoscalar effective mass plot  for
$\beta$\,=\,5.9, $\kappa$\,=\,0.1385.  The $1\times1$ one-state (diamond), 
$2\times2$ two-state (square), $3\times3$ three-state (circle) fits are shown.}
\label{fig:Eff}
\end{figure}

We use the multi-state smearing method \cite{Defhht95} to suppress
excited state contamination. The smearing sources are fits to the
measured wavefunctions of the pseudoscalar ground and excited states
with the following forms,
\begin{eqnarray}
f_{1S}(\mbox{r}) & = & \exp( -\mu_{1S} \mbox{r} ),  \\
f_{2S}(\mbox{r}) & = & ( 1 - \nu_{2S} ) 
\exp( -\mu_{2S} \mbox{r} ).
\end{eqnarray}
For $\beta$\,=\,5.5, 5.7 and 5.9, we use 1S, 2S and local sources, while
for $\beta$\,=\,6.1, only 1S and local sources are used. We choose $2\times2$
two-state fits as our best fits. In order to estimate the systematic
error of excited state contamination, we compare our best fits with
the results from $1\times1$ one-state and $3\times3$ three-state fits.
We find that the difference is less than 1\% for $\beta$\,=\,5.7, 5.9
and about 1-1.5\% for $\beta$\,=\,6.1. (See Figure~\ref{fig:Eff}.)
%

\begin{figure}[tb]
\centerline{\epsfysize=5.0cm \epsffile{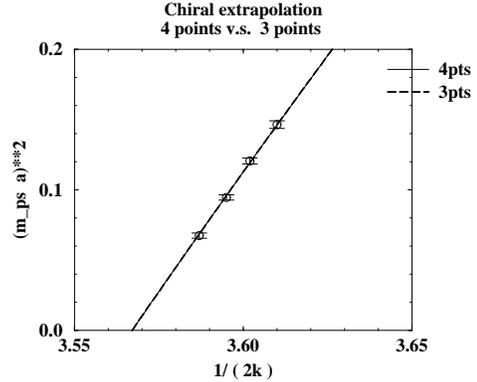}}
\caption{The Chiral extrapolation of $m_{PS}^2$ for $\beta$\,=\,5.9.
The four $\kappa$ fit (solid line) and the three $\kappa$ fit 
(dashed line) are almost indistinguishable.}
\label{fig:chiral}
\end{figure}

As the chiral extrapolation error, we take the difference in the
chiral extrapolation with three $\kappa$'s and four $\kappa$'s. The
results are again less than 1\% for $\beta$\,=\,5.7 and 5.9, and about
3\% for $\beta$\,=\,6.1. (See Figure~\ref{fig:chiral}.)
%

The one loop the renormalization factor which connects
the lattice bare mass with $\overline{MS}$ mass is,
%
%
\begin{equation}
\overline{m}_q(\mu) =
\widetilde{m}_q^{lat}
\left[
1{+}\alpha_{\mbox{v}}(q^{\ast})
( \gamma_0  
( \ln
\widetilde{C}_m{-}\ln( a
\mu)
) 
)
\right].
\label{m0msbar}
\label{eq:oneloop}
\end{equation}
The mean-field improved bare mass $\widetilde{m}_q^{lat}$
is given by $\widetilde{m}_q^{lat} = m_0/(1 - (\pi/3)
\alpha_{\mbox{v}}
 + \dots ~)$
in perturbation theory \cite{LM93}.
$\gamma_0=2/\pi$ is the leading quark mass
anomalous dimension.
$\widetilde{C}_m$ for SW-improved light quarks is 4.72  \cite{Gab91}.

Using Eq.(\ref{eq:oneloop}), we first convert the lattice quark mass to
the $\overline{MS}$ mass at $\mu=\pi/a$ or $1/a$, then run it to the
common scale of 2 GeV. In Eq.(\ref{eq:oneloop}), there is another scale
$q^{\ast}$, which is the scale for the gauge coupling constant. Since
we do not know the two-loop correction, it is not obvious which scale
we should take for $q^{\ast}$. We estimate the size of unknown higher
order corrections to Eq.(\ref{eq:oneloop}) by varying $q^{\ast}$
between $1/a$ and $\pi/a$.
%
%
This procedure is consistent with assuming a coefficient of order
unity for the $\alpha_{\mbox{v}}^2$ term.
Our estimates are 30\%, 13\%, 7\%, 5\%
       for $\beta$ = 5.5, 5.7, 5.9, 6.1.
\begin{table}
\caption{Our results for 
$A$, the coefficient of the linear fit in the $m_{\pi}^2$ 
chiral extrapolation, $\widetilde{m}_q^{lat}$ (in MeV), the 
tadpole improved lattice bare mass, and  
$\overline{m}_q(2\,{\rm GeV})$ (in MeV), 
in the $\overline{MS}$ scheme, 
renormalized at 2 GeV ($q^{\ast} = \pi/a, 1/a$.) . }
\label{tab:mq}
\begin{tabular}{l|cccc}
\hline
\hline
$\beta$&5.5&5.7&5.9&6.1\\
\hline
$A$&5.3(7)&4.1(5)&3.05(7)&2.24(10)\\
$\widetilde{m}_q^{lat}$&4.34(17)&3.9(1)&3.3(1)&3.2(1)\\
$\overline{m}_q(\pi/a)$ & 4.75(19)&4.41(12)&3.90(13)&3.84(18)\\
$\overline{m}_q(1/a)$&6.20(25)&4.89(13)&4.19(14)&4.05(19)\\
\hline
\hline
\end{tabular}
\end{table}
%

\begin{figure}
\centerline{\epsfysize=5.0cm \epsffile{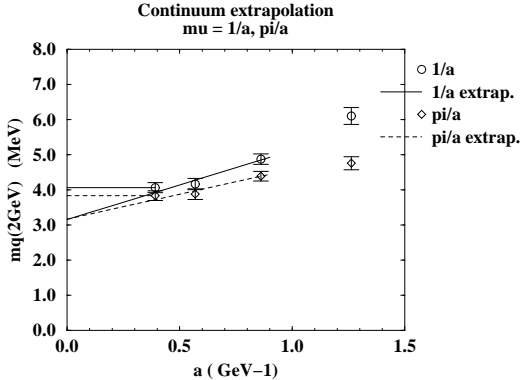}}
\caption{The continuum extrapolation of $\overline{m}_q$.
The upper value is the $\beta$\,=\,6.1 result, and the lower value is the 
naive linear extrapolation of the $\beta$\,=\,5.7, 5.9, 6.1 data.
The data for $\mu=q^{\ast}$=1/a (circle, solid line), $\pi$/a 
(diamond, dashed line) are presented in the same graph.}
\label{fig:Continuum}
\end{figure}

There are both $O( \alpha a )$ and $O( a^2 )$ corrections to the
action, and the continuum extrapolation could change depending on the
relative size of these subleading terms. All we can say is that there
is a systematic downward trend as we approach to the continuum.
Without a theoretical argument to tell us about the $a$-dependence, we
take the $\beta$\,=\,6.1 result as an upper value and take the linearly
extrapolated value using $\beta$\,=\,5.7, 5.9, 6.1 as a lower value.
Our estimate of the continuum extrapolation error is 11\%. (See
Figure~\ref{fig:Continuum}.)
\section{Summary}
In summary, our error estimates are,
\begin{center}
\begin{tabular}{lr}
excited states			& $<$ 1.5\%  \\
chiral extrapolation		& $\sim$ 3\%  \\
perturbative			& 5\%  \\
continuum extrapolation	& 11\%  \\
\hline
\hline
combined			& 17\%  \\
\end{tabular}
\end{center}
The perturbative and $a$ dependent errors are intertwined. We combine them
linearly in the following way. As we saw earlier, the scale
of the coupling constant $q^{\ast}$ is arbitrary. When we discuss the
continuum limit, we therefore perform the extrapolation of the data
for both $q^{\ast}=1/a$ and $\pi/a$ (Figure~\ref{fig:Continuum}).
The outer points so obtained are taken as the limits of the combined
error bar. The remaining errors are much smaller and combined in
quadrature. Our final result for the light quark mass in the
$\overline{MS}$ scheme in the quenched approximation is
\begin{eqnarray}
	\overline{m}_q(\mu=\mbox{2\,GeV};n_f=0) = 3.6 \pm 0.6\,\mbox{MeV}.
\end{eqnarray}
\section{Acknowledgments}
We wish to thank our colleagues in the Fermilab Computing Division. TO
and AXK thank the Fermilab theory group for their kind
hospitality. These calculations were performed on the Fermilab {\sc
acpmaps} supercomputer.

%
\end{document}